\documentclass[12pt]{article}

\usepackage{amssymb}
\usepackage{amsmath}
\usepackage{amscd}
\usepackage{latexsym}
\usepackage{graphicx}

\newcommand{\be}{\begin{equation}}
\newcommand{\ee}{\end{equation}}
\newcommand{\Dlt}{\Delta}
\newcommand{\dlt}{\delta}
\newcommand{\prt}{\partial}
\newcommand{\br}{{\bf r}}

\newcommand{\bt}{\beta}
\newcommand{\vp}{\varphi}

\newcommand{\al}{\alpha}

\newcommand{\gm}{\gamma}
\newcommand{\om}{\omega}
\newcommand{\Om}{\Omega}

\newcommand{\dgr}{\dagger}
\newcommand{\lbd}{\lambda}
\newcommand{\Lbd}{\Lambda}
\newcommand{\cD}{{\cal D}}

\newcommand{\cH}{{\cal H}}

\newcommand{\rgl}{\rangle}
\newcommand{\lgl}{\langle}

\begin{document}

\begin{center}

{\Large{\bf Statistical models of nonequilibrium Bose gases} \\ [5mm]

V.I. Yukalov$^{1}$ and E.P. Yukalova$^{2}$ } \\ [3mm]

{\it
$^1$Bogolubov Laboratory of Theoretical Physics, \\
Joint Institute for Nuclear Research, Dubna 141980, Russia \\ [3mm]

$^2$Laboratory of Information Technologies, \\
Joint Institute for Nuclear Research, Dubna 141980, Russia }

\vskip 2mm

E-mail: yukalov@theor.jinr.ru

\end{center}

\vskip 0.5cm

\begin{abstract}

The idea is advanced that strong perturbations of an initially equilibrium 
Bose-condensed gas lead to the sequence of nonequilibrium states whose order 
is inverse to the sequence of states arising in the process of the Bose-gas 
relaxation from an initial nonequilibrium state. An approach is described 
for constructing statistical models of nonequilibrium Bose gases. The method 
is based on the averaging over heterogenous configurations of a nonequilibrium 
system. A statistical model of grain turbulence is suggested. A simple model
is analyzed consisting of a mixture of two phases, one gauge symmetric and the 
other with broken gauge symmetry.

\end{abstract}

\newpage

\section{Introduction}

The description of nonequilibrium systems is notoriously difficult, since
such systems are usually strongly nonuniform and their dynamical states may 
quickly vary. However, when it is possible to separate in the system dynamics 
a regime exhibiting specific properties during sufficiently long time, longer 
that the time of fast local oscillations, then it may be admissible to average 
over the local fluctuations and to reduce the consideration to an effective 
quasi-stationary system, whose treatment is essentially simpler than that of the 
generic nonequilibrium system. As examples, we can mention the statistical 
models of fully developed vortex turbulence [1-3], heterophase models of 
quasi-equilibrium systems [4,5], and effective averaged models of 
nonequilibrium trapped atoms subject to periodic perturbations [6].  

In the present paper, we describe a general approach for constructing effective
quasi-stationary models for nonequilibrium systems. The approach is applicable 
when a nonequilibrium system exhibits some specific properties during sufficiently 
long time that is longer than the characteristic time of fast oscillations.

The transfer of a system from an equilibrium state to nonequilibrium states
can be done by imposing external perturbations. For example, a system can be 
subject to a time-dependent perturbation potential $\hat{V}(t)$. Then the 
energy injected into the system by this potential is defined as
$$
 E_{inj} \equiv \int_0^t \left | \left \lgl 
\frac{\prt\hat V(t')}{\prt t'} \right \rgl \right | \; dt' \;  .   
$$ 
In an averaged statistical picture, the injected energy acts similarly to 
temperature in equilibrium systems. Therefore, for a nonequilibrium system 
of $N$ atoms, it is possible to introduce the effective temperature
$$
T_{eff} \equiv \frac{E_{inj}}{k_B N} \;   ,
$$
which, for brevity, we may denote just as $T$. 

The outline of the paper is as follows. In Sec. 2, a typical way of creating 
strongly nonequilibrium Bose-condensed gases is described and the 
characteristic parameters of Bose-condensed gases, used for creating such 
strongly nonequilibrium states, are discussed in order to give the feeling 
of the scales involved in experiments with nonequilibrium trapped atoms. 
In Sec. 3, we advance the idea that the procedure of generating a strongly 
nonequilibrium Bose system from an initially equilibrium Bose-Einstein 
condensate follows the same sequence of states, although in the reverse order, 
as the process of relaxation of an initially strongly nonequilibrium Bose 
system equilibrating to its condensed state. In Sec. 4, we describe the 
general method of deriving effective quasi-stationary models for nonequilibrium 
systems. The method is based on averaging over heterogeneous configurations 
arising in the treated nonequilibrium system. The consideration is exemplified 
in Sec. 5 by a statistical model of grain turbulence. In Sec. 6, we study 
a simple statistical model representing a nonequilibrium mixture of two phases
with different symmetries, one gauge symmetric and the other with broken gauge 
symmetry. Section 7 concludes.  

When it does not lead to confusion, we employ the system of units where the 
Planck and Boltzmann constants are set to unity.

\section{Excitation of trapped Bose-Einstein condensates}

Trapped atoms provide a convenient tool for studying strongly nonequilibrium 
states of quantum systems. In the process of excitation, an atomic system 
passes through a sequence of qualitatively different states. The system of 
trapped condensed atoms can be strongly perturbed by an external field, when 
gradually increasing its strength and time of action. 

Trapped Bose atoms in equilibrium at low temperatures form Bose-Einstein
condensate in the ground state. The condensate cloud in a trap enjoys
approximately Thomas-Fermi shape, with well known properties described in the
books [7-9] and reviews [10-19]. In order to break the condensate into pieces, 
it is necessary to impose external perturbations transferring the condensate 
from its ground state to excited states. There are two main ways of imposing 
such external perturbations.

One possibility is to add to the static trapping potential $U({\bf r})$ an
alternating potential $V({\bf r},t)$, so that the total trap potential becomes
\be
\label{1}
 U(\br,t) = U(\br) + V(\br,t) \;  .
\ee
Another way is to modulate the scattering length $a_s(t)$ by means of
Feshbach resonance techniques. Both these ways can be used for strongly
disturbing condensate [19].

Suppose trapped Bose atoms have been cooled down to very low temperatures,
when practically all of them pile down to a Bose-condensed state. And let us 
apply an external modulating perturbation by one of the methods mentioned above. 
First, at weak perturbation, there appear elementary collective excitations, 
that are small deviations from the ground state. Weak perturbations also can 
generate large deviations from the ground state, provided that the modulation 
frequency is in resonance with one of the transition frequencies between 
{\it topological coherent modes} [20]. The latter are defined as the 
eigenfunctions of the stationary nonlinear Schr\"{o}dinger (NLS) equation 
\be
\label{2}
 \left [ -\; \frac{\nabla^2}{2m} + U(\br) + N \Phi_0 | \vp_n(\br)|^2
\right ] \vp_n(\br) = E_n \vp_n(\br) \;  ,
\ee
where $N$ is the number of condensed atoms, assumed to be close to the total 
number of atoms, and
\be
\label{3}
\Phi_0 \equiv 4\pi\; \frac{a_s}{m}
\ee
is the atomic interaction strength, in which $a_s$ is a scattering length
assumed to be positive. The modes are termed topological, since different 
modes have different numbers of zeroes, thus, topologically different atomic 
densities. The modes are coherent, being formed by condensed atoms characterized 
by coherent states.

The known particular example of the topological modes are quantum vortices. If
the external perturbation rotates the atomic cloud, acting as a spoon, then
vortices appear being aligned along the imposed axis of rotation. But when the
trap modulation does not prescribe a fixed rotation axis, then vortices and
antivortices arise in pairs or in larger groups [10,20]. The explicit experimental 
demonstration for the appearance of clusters of vortices and antivortices was 
done in Ref. [21].   

Increasing the strength of the trap modulation generates a variety of coherent
modes, needing no resonance conditions because of the power broadening effect [22].
Among these numerous coherent modes, the basic vortex, with the winding number 
one, is the most energetically stable. For a trap with a transverse, 
$\omega_{\perp}$, and longitudinal, $\omega_z$, frequencies, the vortex energy 
can be written [9] as
\be
\label{4}
 \om_{vor} = \frac{0.9\om_\perp}{(\nu g)^{2/5} } \;
\ln ( 0.8 \nu g ) \;  ,
\ee
where the notation is used for the trap aspect ratio
\be
\label{5}
\nu \equiv \frac{\om_z}{\om_\perp} = \left ( \frac{l_\perp}{l_z}
\right )^2
\ee
and for the effective coupling parameter
\be
\label{6}
 g \equiv 4\pi N \; \frac{a_s}{l_\perp} \; ,
\ee
with $l_{\perp}$ and $l_z$ being the transverse and longitudinal oscillator
lengths, respectively. Due to the large number of atoms $N$, the effective
coupling parameter is large, $g \gg 1$. As is seen, the basic vortex energy
diminishes with the increase of $g$. At the same time, the transition
frequencies of other modes, hence their energies, can be shown [20,22]
to increase as
\be
\label{7}
 \om_{mn} \; \propto \; (\nu g)^{2/5} \qquad ( g \gg 1) \; .
\ee
This makes the basic vortex the most energetically stable mode.

When the trap aspect ratio is not too small, the trap can house many vortices,
whose number can be estimated as
\be
\label{8}
N_{vor} \sim \frac{E_{inj}}{\om_{vor}} \;   ,
\ee
where $E_{inj}$ is the energy injected into the trap by the external pumping. 
The vortices are created due to dynamical instability arising in the moving 
fluid [23-30]. 

Increasing the strength of the pumping, without imposing a rotation axis, 
produces a tangle of vortices, which makes the trapped atomic cloud turbulent
[31-34]. Increasing further either the amplitude of the pumping field or the 
pumping time leads to the appearance of the condensate granulation [35].

The energy per particle, injected into the trap by the external perturbation, 
as is explained in the Introduction, plays the role of an effective temperature
\be
\label{9}
 T_{eff} = \frac{1}{N} \int \rho(\br,t) \left |
\frac{\prt V(\br,t)}{\prt t} \right | \; d\br dt \;  ,
\ee
where $\rho({\bf r}, t)$ is atomic density. In the case of the periodic in time
alternating field $V({\bf r},t) \sim A \cos(\omega t)$, the energy per atom, 
injected during the time interval $[t_1,t_2]$, takes the form 
$T_{eff} \approx \omega (t_1 - t_2)$. This makes it possible to represent the 
crossover lines between different regimes as the relation
\be
\label{10}
A = \frac{T_{eff}}{\om(t_1-t_2)}
\ee
between the amplitude $A$ of the pumping field and the pumping time.

The experimental phase diagram on the amplitude-time $A-t$ plane is described 
in Refs. [33-35], where it is shown that with increasing the injected energy, 
that is proportional to the product $A t$, the system passes through the 
following states: {\it regular superfluid} slightly perturbed by 
a weak external field, {\it vortex superfluid} with several vortices, 
{\it turbulent superfluid} formed by a tangle of many vortices, and 
{\it granular state} with condensate droplets surrounded by uncondensed gas.

To give the reader the feeling of the typical parameters in the experiments
with trapped atomic gases, let us mention the corresponding characteristic
quantities and experimental data.  

The size of an atomic cloud can be found by solving the nonlinear Shr\"{o}dinger 
(NLS) equation and calculating the mean-square lengths in the corresponding 
directions. For a cylindrical trap, taking into account that $g \gg 1$, this 
gives [10,20,22] the mean transverse radius
\be
\label{11}
r_\perp = \frac{(2\nu g)^{1/5} l_\perp}{(2\pi)^{3/10}} =
0.662 (\nu g)^{1/5} l_\perp
\ee
and the mean axial radius
\be
\label{12}
z_0 = \frac{(\nu g)^{1/5} l_z}{(4\pi)^{3/10}\sqrt{\nu}} =
0.468 \; \frac{(\nu g)^{1/5}}{\sqrt{\nu}}\; l_z \; ,
\ee
where the standard notations are used for the effective transverse oscillator
length $l_\perp \equiv\sqrt{\hbar/m \omega_\perp}$ and longitudinal oscillator
length $l_z \equiv\sqrt{\hbar/m \omega_z}$, and where $\nu$ is the trap aspect
ratio. The mean effective cloud radius is
\be
\label{13}
 r_0 \equiv \left ( r^2_\perp z_0 \right )^{1/3} =
0.59\; \frac{(\nu g)^{1/5}}{\nu^{1/6}}\; l_0 \;  ,
\ee
where the average oscillator length is
$$
 l_0 \equiv \sqrt{ \frac{\hbar}{m\om_0} } \; , \qquad
\om_0 = \left ( \om^2_\perp \om_z \right )^{1/3} \; .
$$
As we see, the actual cloud sizes are noticeably larger than the oscillator
lengths because of repulsive atomic interactions. These sizes even can be
essentially larger than the oscillator lengths, when $g \gg 1$.

Knowing the size of the trapped atomic cloud, it is straightforward to find
the effective cloud volume
\be
\label{14}
V_{eff} \equiv \pi r^2_\perp 2 z_0 = 2 \pi r_0^3 =
1.29 \; \frac{(\nu g)^{3/5}}{\sqrt{\nu}} \; l_0^3
\ee
and to estimate the average atomic density in the trap
\be
\label{15}
\rho \equiv \frac{N}{V_{eff}} = 0.775 \;
\frac{\sqrt{\nu} N}{(\nu g)^{3/5} l_0^3 } \;   .
\ee
This shows that, for strongly repulsive atoms, the atomic density $\rho$ can
be much smaller than the density $N / l_0^3$ they would have in the absence 
of repulsive interactions.

Pair atomic interactions are conveniently characterized by the gas parameter
\be
\label{16}
 \gm \equiv \rho^{1/3} a_s = \frac{a_s}{a} \;  ,
\ee
where $a = \rho^{1/3}$ is mean interatomic distance. The gas parameter $\gamma$ 
is usually small for trapped atoms, though can be varied in a wide range
by means of the Feshbach resonance techniques. Because of the large number of
atoms in a trap, the effective coupling parameter $g$ is usually large.

An important quantity, showing whether atoms are in local equilibrium, is the
local equilibration time $t_{loc} \sim m/\hbar\rho a_s$. A perturbed cloud of 
trapped atoms can, as a whole, be strongly nonequilibrium, while, at the same 
time, be locally equilibrium. This happens in the situation, when the
modulation period $t_{mod} \equiv 2 \pi /\omega$ of the alternating modulating
field, with frequency $\omega$, is much longer than the local equilibration
time $t_{loc}$.

In experiments [33-35], strongly nonequilibrium states were generated by
modulating the trapping potential for a trapped cloud of $^{87}$Rb. The cloud 
of $^{87}$Rb atoms, of mass $m = 1.443 \times 10^{-22}$ g and scattering
length $a_s = 0.557 \times 10^{-6}$ cm, has been cooled down to the temperatures
much lower than the Bose-Einstein condensation temperature $T_c = 276$ nK, 
so that the great majority of all $N = 2 \times 10^5$ atoms have been condensed, 
the condensate fraction being $n_0 = 0.7$.

The trap had cylindrical shape, with the parameters
$$
\om_\perp = 1.32 \times 10^3 {\rm s}^{-1} \; , \qquad
\om_z = 1.45 \times 10^2 {\rm s}^{-1} \; ,
$$
$$
l_\perp = 0.74 \times 10^{-4} {\rm cm} \; , \qquad
l_z = 2.25 \times 10^{-4} {\rm cm} \; ,
$$
\be
\label{17}
\om_0 = 0.63 \times 10^{3} {\rm s}^{-1} \; , \qquad
l_0 = 1.08 \times 10^{-4} {\rm cm} \;    ,
\ee
which gives the trap aspect ratio $\nu = 0.11$. The effective coupling parameter
is $g = 1.96 \times 10^4$.

The atomic cloud is characterized by the sizes
$$
r_\perp = 2.27 \times 10^{-4} {\rm cm} \; , \qquad
z_0 = 1.47 \times 10^{-3} {\rm cm} \; ,
$$
\be
\label{18}
r_0 = 0.42 \times 10^{-3} {\rm cm} \; , \qquad V_{eff} = 0.47\times
10^{-9} {\rm cm}^3 \; ,
\ee
which define the effective atomic density $\rho = 0.43 \times 10^{15}$ cm$^{-3}$
and the mean interatomic distance $a = 1.32 \times 10^{-5}$ cm. The gas parameter
is $\gamma = 0.044$.

The trap potential was modulated by an additional alternating potential
$V({\bf r}, t)$ [36,37] oscillating with frequency 
$\omega = 1.26 \times 10^3$ s$^{-1}$, which corresponds to the modulation period
$t_{mod} = 0.5 \times 10^{-2}$ s. The total modulation time $t_{ext}$ was varied
between $0.02$ s and $0.1$ s. The local equilibration time is 
$t_{loc} = 0.57 \times 10^{-3}$ s. Thus, the relations between the characteristic
times is $t_{loc} \ll t_{mod} \ll t_{ext}$.

Because of the high atomic density inside the trap, the {\it in situ} observation
was impossible. Absorption pictures were taken in the time-of-flight setup, after
the times $t_{tof}$ between $0.015$ s and $0.023$ s. Restoring the characteristic 
linear size of granules, corresponding to the experimental situation before the 
free expansion, one gets [35] $l_g \approx 3 \times 10^{-5}$ cm. This is in 
agreement with the theoretical evaluation of the grain size assumed to be of the
order of the coherence length $\xi = 1/\sqrt{4 \pi \rho a_s}$. Thus, the relation 
between the characteristic lengths is $a_s \ll a \sim \xi  \sim l_g \ll l_0 < r_0$.

The excitation of strongly nonuniform states can also be realized by modulating
the scattering length [38,39]. Generally, long modulation times or large exciting 
amplitudes generate the cloud evolution from the appearance of separate vortices 
to tangled vortex configurations, typical of quantum turbulence, and to the 
granular state.   

To compare the parameters in the experiments with trapped $^{87}$Rb atoms with 
the typical parameters of other experiments with trapped atoms, let us consider
the case of $^7$Li in a light trap formed by focused laser beams, as described 
in Refs. [40,41]. The atoms of $^7$Li of mass $m = 1.2 \times 10^{-23}$ g are 
prepared, using Feshbach resonance techniques [42], at the scattering length 
$a_s = 3.2 \times 10^{-8}$ cm. Lowering down temperature, essentially below the 
critical condensation temperature $T_c = 200$ nK, almost all $N = 3 \times 10^5$ 
trapped atoms are Bose-condensed, with the condensate fraction $n_0 = 0.9$.

The trap characteristics are
$$
\om_\perp = 1.48 \times 10^3 {\rm s}^{-1} \; , \qquad
\om_z = 30.4 {\rm s}^{-1} \; ,
$$
$$
l_\perp = 2.5 \times 10^{-4} {\rm cm} \; , \qquad
l_z = 1.7 \times 10^{-3} {\rm cm} \; ,
$$
\be
\label{19}
\om_0 = 4.1 \times 10^{2} {\rm s}^{-1} \; , \qquad
l_0 = 4.7 \times 10^{-4} {\rm cm} \;    ,
\ee
which give the trap aspect ratio $\nu = 0.0207$. This means that the trap is
quasi-one-dimensional. 

The cloud shape corresponds to the sizes
$$
r_\perp = 2.62 \times 10^{-4} {\rm cm} \; , \qquad
z_0 = 0.88 \times 10^{-2} {\rm cm} \; ,
$$
\be
\label{20}
r_0 = 0.84 \times 10^{-3} {\rm cm} \; , \qquad V_{eff} = 0.38\times
10^{-8} {\rm cm}^3 \; ,
\ee
which define the effective density $\rho = 0.79 \times 10^{14}$ cm$^{-3}$ and
the mean interatomic distance $a = 2.33 \times 10^{-5}$ cm. This density is an
order lower than that for $^{87}$Rb. The lower density and the larger spatial
size allows to observe the behavior of the atomic cloud {\it in situ}. The gas 
parameter is $\gamma = 1.37 \times 10^{-3}$. The effective coupling parameter 
is $g = 0.48 \times 10^3$.

The effective scattering length
$$
a_s(t) = a_{BG} \left [ 1 \; - \; \frac{\Dlt B}{B(t)-B_\infty} \right ]
$$
can be modulated by varying the magnetic field
$$
 B(t) = B_0 + \dlt B \cdot \cos(\om t) \;  .
$$
Here $a_{BG}$ is a background scattering length far from the resonance field 
$B_\infty$, and $\Delta B$ is the resonance width.

In the case, when $\delta B$ is much smaller than $B_0$, one can write the
oscillating scattering length in the form
$$
a_s(t) \cong a_s + \dlt a_s \cdot \cos(\om t) \;  ,
$$
in which
$$
a_s \equiv a_{BG}\left ( 1\; - \; \frac{\Dlt B}{B_0-B_\infty}
\right )\; , \qquad
\dlt a_s \equiv a_{BG} \; \frac{\Dlt B\dlt B}{(B_0-B_\infty)^2} \;  .
$$

The amplitude of the scattering-length oscillations corresponds to
$\delta a_s / a_s = 0.2$. The frequency $\omega$ is varied in the range between
$157$ s$^{-1}$ and $314$ s$^{-1}$, which gives the modulation period $t_{mod}$
between $2 \times 10^{-2}$ s and $4 \times 10^{-2}$ s. The local equilibration
time $t_{loc} = 4.5 \times 10^{-3}$ s is much shorter than the modulation time,
$t_{loc} \ll t_{mod}$.

At the beginning, the scattering-length modulation, generates quadrupole mode
excitations. The amplitude of the oscillations as a function of the applied 
frequency, allows one to locate the resonance curve for the quadrupole mode 
excitations, as has been shown in Ref. [39].

The relation between the characteristic lengths, including the scattering length
$a_s$, mean interatomic distance $a$, longitudinal oscillator length $l_z$, 
and the axial radius of the cloud $z_0$, is such that $a_s \ll a \ll l_z < z_0$.

The phase diagram on the amplitude-time $A-t$ plane, observed in the experiments
with a strongly perturbed gas of $^{87}$Rb, is discussed in Refs. [33-35]. And 
the detailed results for a strongly nonequilibrium gas of $^7$Li will be 
published in a separate paper.

\section{Nonequilibrium condensation versus strong perturbation}

In the process of strong perturbation, trapped Bose gases pass through several
nonequilibrium states, such as the {\it vortex state} with a few vortices, 
strong {\it vortex turbulence}, and {\it granular state}. Increasing further 
perturbation should completely destroy the condensate, transferring the whole 
system into a chaotic normal state with no condensate [35]. 

To our understanding, the sequence of dynamic states, appearing in the process 
of strong perturbation of trapped Bose gases, should correspond, although in 
the reverse order, to the sequence of dynamic states, arising in the process 
of equilibration of an initially strongly nonequilibrium Bose system to its 
equilibrium Bose-condensed state.   

The equilibration of weakly interacting Bose systems, from an initial strongly 
nonequilibrium normal state to Bose-condensed state has been studied in a 
number of publications. Levich and Yakhot [43] tried to describe this process 
by a kinetic equation for the occupation number of particles. They found that 
the time of Bose condensation is infinite, and becomes finite only if the 
presence of germs of the Bose-condensed phase at the beginning of the cooling 
process is assumed. Thus, the single {\it kinetic stage} cannot result in 
Bose-Einstein condensation. 

Stoof [44,45] employed a functional approach in the frame of the Keldysh 
formalism for deriving a time-dependent Ginzburg-Landau theory and a 
Fokker-Planck equation for the initial stages of nonequilibrium Bose-Einstein 
condensation. He distinguishes three stages of the nonequilibrium condensation. 
After the Bose gas is quenched into the critical region of the phase transition, 
at the first stage, it can be described by the quantum Boltzmann equation. 
However, such a kinetic equation cannot describe the buildup of coherence in 
the gas and therefore does not lead to a microscopic occupation of the 
single-particle state. In other words, incoherent collisions, governing the 
Boltzmann kinetic equation, cannot lead to Bose-Einstein condensation [43]. 
To achieve this, a second stage is needed, in which the gas develops the 
instability toward Bose condensation and then coherently populates the ground 
state by a depletion of the low-lying excited states. The assumed instability, 
developing in the coherent stage, is analogous to a dynamic phase transition.  
After the coherent stage, the gas acquires a highly nonequilibrium energy 
distribution, and equilibrates during the third and final stage. This last 
stage is again of kinetic nature and can be described by the appropriate 
quantum Boltzman equation for the Bogolubov quasiparticles of the Bose-condensed 
gas. In that way, according to Stoof, there are three stages of nonequilibrium 
condensation: {\it kinetic stage}, {\it coherent stage}, and {\it relaxation stage} 
that also is of kinetic type.           
    
The kinetic Boltzmann equation was solved numerically by Snoke and Wolfe [46]
and by Semikoz and Tkachev [47]. In the latter paper, the authors acknowledge 
the existence of three stages, as in the Stoof picture, but treat only two 
kinetic stages. At the first stage, the condensate is absent, but there is a 
nonzero inflow of particles towards the zero momentum state. In the framework 
of the kinetic equation, there is no condensate at all times if there was no 
condensate initially. Therefore one has to add a seed condensate by hands when 
switching from the first kinetic stage to the final kinetic stage. In this 
approach, the intermediate coherent stage, where the condensate actually emerges 
by a kind of a phase transition, is omitted.

The process of Bose condensation has also been considered by numerical solutions 
of the NLS equation assuming the presence of the solution random phase [48] or
of external random noise [49]. The condensation stages were not clearly 
distinguished.    

The occurrence of several stages in nonequilibrium Bose condensation of weakly 
interacting Bose gas have been emphasized by Kagan and Svistunov [50,51] and 
Berloff and Svistunov [52]. Strongly nonequilibrium Bose gas, after a very short 
time develops a highly chaotic state, where kinetic energy is much larger than
the interaction energy. Because of the small nonlinearity, the system can be 
treated as a collection of almost independent modes with random phases. The 
smallness of correlations between the modes implies the absence of any order. 
This chaotic regime of normal (non-superfluid) gas is termed wave turbulence or 
weak turbulence. Assuming that the low-energy modes are macroscopically occupied 
(so that the occupation numbers are much greater than one), it is possible to 
represent the system by a coherent field with random phase. The modes propagate 
from relatively high to lower energies, and at some time the wave turbulence 
transforms into a regime where short-range coherence starts appearing. This is 
the regime of strong turbulence that cannot be characterized by quasi-independent 
modes. After this intermediate stage, the regime of superfluid turbulence arises, 
where numerous tangled vortices form a {\it random tangle}. This could also be 
called the vortex turbulence. The next stage is the process of relaxation of the 
vortex turbulence to an equilibrium state in a macroscopically long time. In this 
way, one can distinguish four stages: {\it wave turbulence}, {\it strong turbulence},
{\it vortex turbulence}, and {\it relaxation stage}. Since from the very 
beginning, one assumes that the low-energy modes are macroscopically occupied,
the whole process becomes a crossover, with a continuous growth of the low-energy
occupation numbers, the lowest of which represents Bose condensate. The 
intermediate regime of strong turbulence is equivalent, in the Stoof picture,
to his coherent stage. The difference is that Stoof suggests that during this 
regime a kind of phase transition occurs, when the germs of Bose condensate 
suddenly appear. While in the Svistunov et all picture, the low-energy modes 
are assumed to exist already in the regime of wave turbulence, just starting 
fast growing in the intermediate strong turbulence stage, so that there is 
not a phase transition but a sharp crossover. It has been mentioned [53] that 
the dynamics of the Bose-Einstein condensation is similar to the collapse dynamics
of a self-gravitating gas. 

The dynamics of the condensation of a weakly interacting Bose gas in a trap 
is analogous to that of the homogeneous gas [54] and can be characterized by
a quantum kinetic equation, where the arising condensate comes from the bath
of uncondensed atoms [55-57].          

Zakharov and Nazarenko [58] distinguish four regimes in the dynamics of 
Bose-Einstein condensation. The first is the kinetic stage, when the system is 
forced into a nonequilibrium state corresponding to weak turbulence. The kinetic 
regime transfers into a strong-turbulence state, where the kinetic description 
breaks down. The second stage is the vortex turbulence, where there appear a 
number of vortices forming a chaotic tangle. In the third stage, the system is 
filled by a well developed condensate with just a few vortices. And the final 
stage corresponds to the relaxation to the equilibrium state. The authors [58]
concentrated their attention on the strong-condensate regime containing a 
small number of vortices. They used the NLS equation complimented by a term
describing forcing and dissipation.   

The condensate dynamics from a nonequilibrium initial state have been studied 
experimentally [59-62], observing the simultaneous population growth and the
development of the phase coherence. The regime of vortex turbulence was 
investigated in experiments [31-35,63] and reviewed in Refs. [35,64-67]. 
 
The regime of vortex turbulence, developing in the process of the nonequilibrium 
Bose-Einstein condensation, is the manifestation of the Kibble-Zurek mechanism 
[68-70] that has been observed [71] in the condensation of trapped $^{87}Rb$ atoms. 
In this picture [68-71], the vortex turbulence is preceded by the formation of
a nonuniform structure composed of the coherent germs, called by Kibble [68] 
"cells", or "protodomains", of the condensed phase inside the cloud of 
uncondensed atoms. The order parameters of different cells are random, so that 
there is no coherence between different protodomains. Such domains move in space 
and can fuse with each other. According to Kibble [68], it is the cell fusion 
that creates vortices. 

In order to clearly distinguish these nonequilibrium coherent cells, or 
protodomains, from the static domains occurring in ferromagnets, we shall call 
these nonequilibrium germs the {\it grains}. Since their sizes are intermediate
between the atomic interaction length and the system size, such mesoscopic germs 
of one phase inside another are analogous to the heterophase fluctuations [4,5].
These coherent grains appear for repulsive atomic interactions, and should not 
be confused with the bound droplet dew forming in a Bose gas with attractive 
interactions [72]. Because the shapes and spatial locations of the grains are 
random, the corresponding strongly nonequilibrium phenomenon can be termed 
{\it grain turbulence}. The regime of grain turbulence happens between the 
stages of wave turbulence and vortex turbulence, hence, corresponding to the 
stage of strong turbulence. This is the most difficult regime for theoretical 
description, being strongly nonequilibrium and produced by the chaotic motion of
the grains of random shapes, which are randomly distributed in space. 
 
To be more concrete, let us characterize the specific dynamic stages of 
nonequilibrium Bose systems. These stages are connected with the typical
length and time scales of statistical systems [4,5,12]. For dilute gases, the 
shortest is the interaction length $r_{int}$ that is much smaller than the 
scattering length $a_s$. Other lengths are: the mean interatomic distance $a$,
correlation length $\xi$, and the mean free path $\lambda$, for which we have
\be
\label{21}
 a \sim \frac{1}{\rho^{1/3}} \; , \qquad \xi \sim \frac{\hbar}{ms} \;  ,
\qquad \lbd \sim \frac{1}{\rho a_s^2} \; ,
\ee
where $\rho$ is the mean atomic density and 
$$
s \sim \frac{\hbar}{m}\; \sqrt{4\pi\rho a_s}
$$
is sound velocity. For Bose gases, the typical relation between the lengths is
$$
 a_s < a < \xi < \lbd \;  .
$$

These lengths are connected with the characteristic velocities: the scattering
velocity $v_s$, kinetic velocity $v_a$, and the sound velocity $s$, so that
\be
\label{22}
  v_s \sim \frac{\hbar}{ma_s} \; , \qquad v_a \sim \frac{\hbar}{ma} \;  ,
\qquad s \sim \frac{\hbar}{m\xi} \;  ,
\ee
with the typical relation
$$
s < v_a < v_s \;   .
$$
The kinetic velocity shows the typical velocity of atoms between collisions, 
while the scattering velocity corresponds to the velocity of atoms in the 
process of their collisions. 
  
The characteristic time scales, respectively, are as follows. The 
{\it interaction time}
\be
\label{23}
t_{int} \sim \frac{a_s}{v_s} \sim \frac{ma_s^2}{\hbar} 
\ee
is the time of atomic interactions. The {\it local-equilibration time}
\be
\label{24}
t_{loc} \sim \frac{\lbd}{v_s} \sim \frac{m}{\hbar\rho a_s}
\ee
is the time during which there develops local equilibrium and there can 
arise correlated regions in space. The {\it heterophase time}
\be
\label{25}
t_{het} \sim \frac{\xi}{a_s}\; t_{loc} \sim \frac{\lbd}{s} \sim \frac{1}{\rho a_s^2 s}
\ee
is the lifetime during which there can exist well correlated regions in space, 
corresponding to protodomaion, or grains. The longer is the relaxation time 
$t_{rel}$, after which the system gradually relaxes to its equilibrium state 
during the equilibration time $t_{equ}$.       

The intervals between the above times define the corresponding dynamic 
stages through which a Bose system passes, being initially prepared in a 
strongly nonequilibrium state by quickly quenching it to the conditions, where
the Bose condensate should occur. The first is the {\it interaction stage},
or {\it dynamic stage},
\be
\label{26}
  0 < t < t_{int} \qquad (interaction \; stage) \; ,
\ee
during which atoms interact with each other a few times. At this short initial 
stage, statistical description is not yet applicable. 

The second is the {\it kinetic stage}, 
\be
\label{27}
 t_{int} < t < t_{loc} \qquad (kinetic \; stage) \;  ,
\ee
where the system can be described by a kinetic equation. This stage corresponds
to the regime of {\it wave turbulence}, or {\it weak turbulence}, when
the kinetic energy is much larger than the interaction energy, so that the 
system can be represented as a collection of almost independent modes of small 
amplitudes. The mode independence is due to their spatial phases being random.
Thus, the regime of wave turbulence is characterized by three features:
{\it large kinetic energy}, {\it modes of small amplitude}, and {\it random 
spatial phases}. There is yet no condensate at this stage. 

The third is the {\it heterophase stage},
\be
\label{28}
  t_{loc} < t < t_{het} \qquad (heterophase \; stage) \;  ,
\ee
when the kinetic description becomes not applicable, since atomic interactions 
and correlations start playing an important role, as a result of which the 
mode amplitudes grow, and there appear mesoscopic well correlated regions of
locally condensed atoms. The regions are mesoscopic, having the typical sizes 
of the correlation length $\xi$ that is between the scattering length $a_s$ 
and the system size. This is the regime of {\it strong turbulence}, or 
{\it grain turbulence}. The atoms inside each grain are well correlated, 
forming a condensed droplet, but different grains are not necessarily correlated 
with each other and possess different random phases. The condensed grains are 
surrounded by a gas of normal uncondensed atoms. 

The fourth is the {\it hydrodynamic stage},
\be
\label{29}
  t_{het} < t < t_{rel} \qquad (hydrodynamic \; stage) \;  ,
\ee
when the mesoscopic condensed grains fuse, forming quantum vortices, according
to the Kibble-Zurek mechanism [68-71]. A tangle of numerous random vortices
arises, creating the regime of quantum {\it vortex turbulence}. This regime can 
be described by superfluid kinetic equations [73] and superfluid hydrodynamic 
equations [74,75]. The hydrodynamic stage lasts till the relaxation time 
$t_{rel}$ that depends on the system parameters and geometry [76,77].

After the relaxation time $t_{rel}$, the vortices decay by mutual recombination 
and phonon emission. For some time, a few vortices survive, but then the system 
tends to its equilibrium Bose-condensed state. This happens at the relaxation 
stage,
\be
\label{30}
 t_{rel} < t < t_{equ} \qquad (relaxation \; stage) \;   ,
\ee
where $t_{equ}$ is the equilibration time [76]. 

To illustrate the values of the characteristic parameters discussed above, 
let us consider the experiments [33-35] with trapped $^{87}$Rb atoms. Then we
have the scattering length $a_s = 0.557 \times 10^{-6}$ cm, mean interatomic 
distance $a = 1.32 \times 10^{-5}$ cm, correlation length 
$\xi \sim 1.823 \times 10^{-5}$ cm, mean free path 
$\lambda \sim 0.75 \times 10^{-2}$ cm, sound velocity $s \sim 0.401$ cm/s, 
kinetic velocity $v_a \sim 0.554$ cm/s, and the scattering velocity 
$v_s \sim 13.12$ cm/s. This defines the characteristic times: the interaction
time $t_{int} \sim 4.245 \times 10^{-8}$ s, local-equilibration time 
$t_{loc} \sim 0.572 \times 10^{-3}$ s, and the heterophase time
$t_{het} \sim 1.87 \times 10^{-2}$ s. The interaction time, as is expected,
is very short, so that atoms quickly pass into the kinetic stage of wave 
turbulence which lasts around $10^{-3}$ s. The duration of the heterophase 
correlated stage is $1.8 \times 10^{-2}$ s, after which the regime of vortex 
turbulence comes into play.    
 
For other setups, the typical parameters can be rather different. For instance,
in the case of $^7$Li, as in experiments [40,41], we have the scattering length
$a_s = 3.2 \times 10^{-8}$ cm, mean interatomic distance 
$a = 2.33 \times 10^{-5}$ cm, correlation length $\xi \sim 1.775 \times 10^{-4}$ cm,
mean free path $\lambda \sim 12.36$ cm, sound velocity $s \sim 0.495$ cm/s,
kinetic velocity $v_a \sim 3.772$ cm/s, and the scattering velocity
$v_s \sim 2.746 \times 10^3$ cm/s. From here, we find the interaction time
$t_{int} \sim 1.165 \times 10^{-11}$ s, local-equilibrium time 
$t_{loc} \sim 4.501 \times 10^{-3}$ s, and the heterophase time 
$t_{het} \sim 24.97$ s. The interaction time is again very short, and atoms 
quickly pass to the kinetic stage of the wave turbulence. But the heterophase
correlated stage is quite long, of order of ten seconds. Hence, more time is 
needed for the development of the vortex turbulence, if any.     

The above description corresponds to the relaxation of a Bose gas that 
initially is prepared in a strongly nonequilibrium normal state, after which 
it tends to its equilibrium condensed state. We suggest that when an initially 
equilibrium condensed Bose gas is subject to strong perturbations, it passes
through the same dynamic stages, although in the reverse order, with the 
increasing amount of the energy pumped into the system. Thus, an initially 
condensed gas, being perturbed, first goes into a nonequilibrium state, where 
just a few vortices can arise. The next is the hydrodynamic stage, where the 
vortex turbulence develops. After this, the heterophase correlated stage 
should occur, where the grain turbulence takes place. Finally, when all 
condensate is destroyed, the regime of wave turbulence has to come. The 
sequence of these stages (except the last one) for a strongly perturbed Bose 
gas of $^{87}$Rb atoms has been observed in experiments [33-35]. The total 
destruction of the condensate requires very strong perturbations that have not 
yet been reached in experiments.    

The dynamic transitions between the different stages described above are,
of course, not absolutely sharp, as would be phase transitions in equilibrium 
systems, but they are rather gradual crossovers.

\section{Averaging over heterogeneous configurations}

Although the dynamic transitions between the nonequilibrium stages are 
crossovers, but inside the corresponding temporal intervals the system
enjoys qualitatively well defined features. This means that it would be 
admissible to introduce quasi-stationary states as those of a system 
averaged over the appropriate time interval. This can be done as follows.  

Let us assume that a nonequilibrium system can live quite long time, being 
supported by an external perturbation with an alternating modulation potential. 
If the local equilibration time is much shorter than the modulation period, 
then the system can be treated as quasi-equilibrium. The difficulty of 
theoretical description of such a system is in its nonuniformity, with 
randomly arising spatio-temporal fluctuations. When the external pumping 
potential does not impose a spatial symmetry, then the nonuniformities are
randomly distributed at each snapshot and do not form any ordered structure. 
Their positions are also random in repeated experiments. The nonuniformities
are often mesoscopic in space, such that their typical sizes are between the 
interaction radius and the system length. In addition, they are usually of 
multiscale nature, with random sizes and shapes in a dense manifold. Such 
nonuniformities are termed {\it heterogeneous}. 

It is possible to show [78] that there exists a mapping between the states of 
an atomic cloud, subject to an alternating modulation, and the states of an 
atomic system in a random spatial external potential. This analogy allows for
the estimation of typical parameters characterizing the process of 
nonequilibrium generation. Assume that such a matter has been created. How 
would it be possible to develop a statistical description of this matter?

Let us consider a snapshot of a heterogeneous matter consisting of the regions 
of two types, whose typical properties can be characterized by some quantities
playing the role of local order parameters or order indices [79]. For instance, 
the role of the order parameters can be played by densities or some local 
atomic configurations. Such regions, with different order parameters, are
analogous to local thermodynamic phases [80,81]. 
  
The spatial separation of the nonuniformities in a system can be described by 
employing the Gibbs theory of quasi-equilibrium systems [80,81]. The total 
system volume can be treated as a union
\be
\label{31}
\mathbb{V} =    \mathbb{V}_1 \bigcup \mathbb{V}_2
\ee
of two parts corresponding to two different spatial regions separated by an 
equimolecular surface. The convenience of using the equimolecular surface is 
that it allows for the additive representation of observable quantities. For 
instance, the system volume and the number of particles are written as the sums
\be
\label{32}
 V = V_1 + V_2 \; , \qquad N = N_1 + N_2 \;  .
\ee
Each subvolume is mathematically characterized by the manifold indicator function
\begin{eqnarray}
\label{33}
\xi_\nu(\br) = \left \{ \begin{array}{ll}
1 , & ~~ \br \in \mathbb{V}_\nu \\
0 , & ~~ \br \not\in \mathbb{V}_\nu
\end{array} \right. \; ,
\end{eqnarray}
where $\nu = 1, 2$ enumerates the local phases, that is, the regions with 
different properties.  

The quasi-equilibrium ensemble, characterizing the system, under a given spatial
distribution of different regions, is the pair $\{\mathcal{H}, \hat{\rho}(\xi)\}$ 
of the space of microstates and a statistical operator. The space of microstates
is the fiber space
\be
\label{34}
\cH \equiv \cH_1 \bigotimes \cH_2
\ee
of weighted Hilbert spaces corresponding to the states typical of a given region 
(phase). The statistical operator is normalized, so that
\be
\label{35}
{\rm Tr} \int \hat\rho(\xi)\; \cD \xi = 1\;   ,
\ee
where the trace is taken over the Hamiltonian degrees of freedom and the 
functional integration is over the manifold indicator functions [4]. This 
functional integration characterizes the random distribution of the shapes, sizes,
and locations of different regions. 

To correctly define the statistical operator, one has to consider a representative
statistical ensemble taking into account all constraints uniquely describing the
system. In addition to the normalization condition, the standard constraint is the
definition of the internal energy
\be
\label{36}
  E =  {\rm Tr} \int \hat\rho(\xi) \hat H(\xi)\; \cD \xi
\ee
as the average of the energy operator $\hat{H}(\xi)$. Other constraints can be
represented as the statistical averages
\be
\label{37}
C_i = {\rm Tr} \int \hat\rho(\xi) \hat C_i(\xi)\; \cD \xi
\ee
of the given constraint operators $\hat{C}_i(\xi)$. The statistical operator
is defined as a minimizer of the information functional
$$
I[ \hat\rho(\xi) ] =
{\rm Tr} \int \hat\rho(\xi) \ln \hat\rho(\xi) \; \cD \xi \; + \;
\lbd_0 \left [ {\rm Tr} \int \hat\rho(\xi)\; \cD \xi - 1 \right ] \; +
$$
\be
\label{38}
+ \; \bt \left [ {\rm Tr} \int \hat\rho(\xi) \hat H(\xi)\; \cD \xi -
E \right ] \; + \; \sum_i \lbd_i \left [
{\rm Tr} \int \hat\rho(\xi) \hat C_i(\xi)\; \cD \xi - C_i \right ]\;,
\ee
in which the first term is the Shannon information and $\lambda_0$, $\beta$, and
$\lambda_i$ are the Lagrange multipliers guaranteeing the validity of the imposed
constraints. This principle of minimal information yields the statistical operator
\be
\label{39}
 \hat\rho(\xi) = \frac{1}{Z} \; \exp \{ - \bt H(\xi) \} \;  ,
\ee
with the grand Hamiltonian
\be
\label{40}
 H(\xi) = \hat H(\xi) - \sum_i \mu_i \hat C_i(\xi) \;  ,
\ee
where $\mu_i = - \lambda_i T$ and $T = 1/\beta$ is effective temperature. The 
inverse normalization factor
$$
 Z =   {\rm Tr} \int \exp\{ -\bt H(\xi) \} \; \cD \xi
$$
is the partition function.

The important point is the introduction of the effective Hamiltonian
$\widetilde{H}$ by the equality
\be
\label{41}
 \int \exp\{ -\bt H(\xi) \} \; \cD \xi =
\exp(-\bt\widetilde H ) \;  .
\ee
Then the partition function takes the simple form
\be
\label{42}
  Z =  {\rm Tr} e^{-\bt\widetilde H} \; ,
\ee
containing only the trace over the Hamiltonian degrees of freedom. The effective
temperature $T$, generally, can include the standard thermal noise and the energy 
injected into the system, as is explained in the Introduction. 

The geometric weights
\be
\label{43}
w_\nu = \int \xi_\nu(\br) \; \cD \xi
\ee
define the probabilities of the related phases, by construction, satisfying the
normalization condition
\be
\label{44}
 w_1 + w_2 = 1 \qquad (0 \leq w_\nu \leq 1 ) \;  .
\ee
These probabilities are found from the minimization of the grand potential
\be
\label{45}
\Om = - T \ln {\rm Tr}  e^{-\bt\widetilde H} \;  .
\ee
The techniques of functional integration over the manifold indicator functions
are thoroughly explained in Ref. [4]. Below, we shall employ these techniques
omitting, for brevity, intermediate calculations.

\section{Statistical model of grain turbulence}

As has been mentioned above, the regime of grain turbulence is one of the most
difficult for theoretical description, if one wishes to consider the details
of its nonequilibrium and nonuniform spatio-temporal behavior. However, if one
is interested in its average statistical properties, it is possible to invoke
the approach of the previous section. This is admissible, since the 
local-equilibrium time $t_{loc}$ is much shorter than the lifetime of this regime 
$t_{het}$. Here we advance a statistical model that can grasp the main average 
features of the heterophase granular state. 

The granular state can be treated as a heterophase mixture of Bose-condensed 
grains immersed into the cloud consisting of normal (non-condensed) atoms.   
Starting with the local-equilibrium Gibbs ensemble [80,81], we average over all
random heterogeneous configurations, as is sketched above, with the related
mathematics thoroughly expounded in reviews [4,5,82]. The resulting effective
Hamiltonian of the mixture becomes the sum of two Hamiltonian replicas
\be
\label{46}   
\widetilde H = H_1 \bigoplus H_2 \;   .
\ee
The Hamiltonian is defined on the fiber space (34). The term $H_1$ acts on the 
space of microstates ${\cal H}_1$ with broken global gauge symmetry $U(1)$, 
corresponding to the Bose-condensed phase [83], while the term $H_2$, on the 
space of microstates ${\cal H}_2$ with preserved gauge symmetry. 

The Bose-condensed phase is characterized by a representative ensemble [84-87], 
with the grand Hamiltonian
\be
\label{47}
H_1 = \hat H_1 -\mu_0 N_0 - \mu_1 \hat N_1 - \hat\Lbd \;   .
\ee   
Here the energy operator is
$$
\hat H_1 =  w_1 \int \hat\psi^\dgr(\br) \left [ -\; \frac{\nabla^2}{2m} + 
U(\br) \right ] \hat\psi(\br) \; d\br \; +
$$
\be
\label{48}
 + \; \frac{w_1^2}{2} \int \hat\psi^\dgr(\br) \hat\psi^\dgr(\br') \Phi(\br-\br')
\hat\psi(\br') \hat\psi(\br) \; d\br d\br ' \; ,
\ee
in which $U(\bf r)$ is an external potential, $\Phi(\bf r)$ is the 
interaction potential, and the Bose field operators are shifted, according 
to Bogolubov [88,89], thus, breaking the gauge symmetry,
\be
\label{49}
 \hat\psi(\br) = \eta(\br) + \psi_1(\br) \;  .
\ee
In this shift, $\eta(\bf r)$ is the condensate wave function and 
$\psi_1(\bf r)$ is the field operator of uncondensed atoms, these terms being 
orthogonal to each other,
\be
\label{50}
\int \eta^*(\br) \psi_1(\br) \; d\br = 0 \;   ,
\ee
which excludes the double counting. The Lagrange multipliers $\mu_0$ and 
$\mu_1$ guarantee the validity of the normalization conditions for the number
of condensed atoms
\be
\label{51}
N_0 = w_1 \int | \eta(\br)|^2 d\br    
\ee
and for that of uncondensed atoms
\be
\label{52}
 N_1 = w_1 \int \lgl  \psi_1^\dgr(\br)\psi_1(\br) \rgl d\br \; .
\ee
The last operator $\hat{\Lambda}$ in Eq. (47) is the Lagrange term preserving
the condition
\be
\label{53}
\lgl  \psi_1(\br) \rgl = 0
\ee
defining the condensate wave function as an order parameter
\be
\label{54}
 \eta(\br) = \lgl  \hat\psi(\br) \rgl \;  .
\ee
  
The grand Hamiltonian of the normal (gauge-symmetric) phase is
\be
\label{55}
 H_2 = \hat H_2 - \mu_2 \hat N_2 \;  .
\ee
Here the energy operator is
$$
\hat H_2 =  w_2 \int \psi_2^\dgr(\br) \left [ -\; \frac{\nabla^2}{2m} + 
U(\br) \right ] \psi_2(\br) \; d\br \; +
$$
\be
\label{56}
+ \; \frac{w_2^2}{2} \int \psi_2^\dgr(\br) \psi_2^\dgr(\br') \Phi(\br-\br')
\psi_2(\br') \psi_2(\br) \; d\br d\br ' \;   .
\ee
The Lagrange multiplier $\mu_2$ guarantees the validity of the normalization
for the number of atoms in the normal phase,
\be
\label{57}
N_2 = w_2 \int \lgl  \psi_2^\dgr(\br)\psi_2(\br) \rgl d\br \;    .
\ee

The total number of atoms in the system is
\be
\label{58}
 N = N_0 + N_1 + N_2 \;  .
\ee
The corresponding atomic fractions
\be
\label{59}
n_0 \equiv \frac{N_0}{N} \; , \qquad n_1 \equiv \frac{N_1}{N} \; , \qquad
n_2 \equiv \frac{N_2}{N} 
\ee
satisfy the normalization
\be
\label{60}
 n_0 + n_1 + n_2 = 1 \;  .
\ee
The condition of heterophase quasi-equilibrium [4,5,82] leads to the relation
\be
\label{61}
  \mu_0 n_0 + \mu_1 n_1 = (n_0 + n_1) \mu_2  
\ee
between the Lagrange multipliers. 

The geometric weights of the phases are defined as the minimizers of the grand
potential. This, with the notation
\be
\label{62}
 w_1 \equiv w \; , \qquad w_2 \equiv 1 - w \;  ,
\ee
implies the conditions
\be
\label{63}
\frac{\prt \Om}{\prt w} = 0 \; , \qquad \frac{\prt^2\Om}{\prt w^2} > 0 \; .
\ee
The first of Eqs. (63), with the use of the notations
$$
K_1 \equiv \int \lgl \hat\psi(\br) \left [ -\; \frac{\nabla^2}{2m} + 
U(\br) \right ] \hat\psi(\br) \rgl \; d\br \; - \; 
\mu_0 \int | \eta(\br)|^2 d\br \; - \; 
\mu_1 \int \lgl \psi_1^\dgr(\br)\psi_1(\br) \rgl \; d\br \; ,
$$
$$
K_2 \equiv \int \lgl \psi_2^\dgr(\br) \left [ -\; \frac{\nabla^2}{2m} + 
U(\br) \right ] \psi_2(\br) \rgl \; d\br \; - \; 
\mu_2 \int \lgl \psi_2^\dgr(\br)\psi_2(\br) \rgl \; d\br \; ,
$$
$$
\Phi_1 \equiv \int \lgl \hat\psi^\dgr(\br) \hat\psi^\dgr(\br') \Phi(\br-\br')
\hat\psi(\br') \hat\psi(\br) \rgl \; d\br d\br' \; , 
$$
$$
\Phi_2 \equiv \int \lgl \psi_2^\dgr(\br) \psi_2^\dgr(\br') \Phi(\br-\br')
\psi_2(\br') \psi_2(\br) \rgl \; d\br d\br' \;   ,
$$ 
yields the equation
\be
\label{64}
 w = \frac{\Phi_2 + K_2 - K_1}{\Phi_1 + \Phi_2}  
\ee
for the weight of the Bose-condensed phase.

The second of Eqs. (63) is the stability requirement imposing the necessary
condition
\be
\label{65}
  \Phi_1 + \Phi_2 > 0 \; .
\ee
This tells us that the effective atomic interactions must be repulsive in
order that the heterophase granular mixture would be stable. 

It is worth emphasizing the necessity of nonzero atomic interactions for 
the Bose-condensed system to be stable. The ideal Bose gas is a pathological 
object that cannot form a stable Bose-Einstein condensate because of 
thermodynamically anomalous particle fluctuations and diverging 
compressibility [12,18,19,77,90,91], although non-condensed ideal Bose gas 
can be stable [92]. But in the ideal Bose gas, there can be neither vortices, 
nor vortex turbulence, nor grain turbulence, whose existence is due to the 
nonlinearity caused by atomic interactions.

\section{Mesoscopic mixture of two phases}

Dealing with the above expressions describing a mesoscopic mixture is a rather
involved problem requiring the use of nontrivial approximations. Meanwhile, 
in order to illustrate how the above approach works, we consider a simplified 
model that can exhibit the mesoscopic coexistence of two phases, one with broken 
gauge symmetry and another with the conserved gauge symmetry. The choice of 
this model is done keeping in mind that Bose-Einstein condensation is necessarily 
accompanied by the global gauge symmetry breaking [83], while the normal, 
non-condensed phase conserves this symmetry. The other advantage of the chosen 
model is its simplicity allowing for a straightforward demonstration of the 
applicability of the suggested approach. The phase with conserved gauge symmetry 
is called disordered and that with the broken symmetry, ordered. The regions 
of the ordered phase, surrounded by the disordered phase, are mesoscopic in the 
sense that their average sizes are larger than the interaction radius, but much 
shorter than the system linear size. This model can be considered as a cartoon 
of a system consisting of Bose-condensed droplets surrounded by normal uncondensed 
gas. 

One should not confuse the considered system with mesoscopic thermal disorder 
with a system in a random potential that induces microscopic quenched disorder [93]. 
The latter system can be equilibrium, while the former is only quasi-equilibrium.   

Let us consider an insulating optical lattice, where atoms can occupy several
energy levels $E_n$. Atoms are assumed to interact through long-range forces, such
as spinor or dipole forces [94]. We start with the standard Hamiltonian in term 
of the field operators $\psi({\bf r})$ that can be expanded over atomic orbitals,
\be
\label{66}
  \psi(\br) = \sum_{nj} c_{nj} \vp_{nj}(\br) \; ,
\ee
where the index $n$ labels energy levels $E_n$ and $j$ enumerates lattice cites.
Keeping in mind unity filling factor, we impose the no-double-occupancy condition
\be
\label{67}
 \sum_n c_{nj}^\dgr c_{nj} = 1 \; , \qquad c_{mj} c_{nj} = 0 \;  .
\ee
Taking account of only two lowest energy levels makes it possible to introduce
the pseudospin operators $S_j^\alpha$  by means of the transformation
$$
c_{1j}^\dgr c_{1j} = \frac{1}{2} + S_j^x \; , \qquad
c_{2j}^\dgr c_{2j} = \frac{1}{2} - S_j^x \; ,
$$
\be
\label{68}
 c_{1j}^\dgr c_{2j} =  S_j^z - i S_j^y \; , \qquad
c_{2j}^\dgr c_{1j} =  S_j^z + i S_j^y \;.
\ee

Assume that the system consists of the random mixture of two different phases.
One phase has broken gauge symmetry and the other is gauge symmetric. By 
accomplishing the averaging over the manifold indicator functions [4], we come 
to the effective Hamiltonian
$$
\widetilde H = H_1 \bigoplus H_2 \; ,
$$
\be
\label{69}
 H_\nu = w_\nu N E_0 + \frac{w_\nu^2}{2} \sum_{i\neq j} A_{ij} -
w_\nu \Om_0 \sum_j S_j^x + w_\nu^2 \sum_{i\neq j} \left ( B_{ij} S_i^x S_j^x
- I_{ij} S_i^z S_j^z \right ) \;  ,
\ee
in which $A_{ij}, B_{ij}, I_{ij}$ are the matrix elements of the atomic
interaction potential [95] and
\be
\label{70}
 E_0 \equiv \frac{1}{2} ( E_1 + E_2 ) \; , \qquad
\Om_0 \equiv E_2 - E_1 \;  .
\ee

This Hamiltonian is invariant under the gauge transformation
\be
\label{71}
 S_j^z \longrightarrow e^{i\al} S_j^z \;  ,
\ee
in which $\alpha$ is a real number. The pseudospin average, depending on the
considered phase, can be calculated in two ways,
\be
\label{72}
 \lgl S_j^z \rgl_\nu \equiv {\rm Tr}_{\cH_\nu} \hat\rho_\nu S_j^z
\qquad (\nu = 1,2) \;  ,
\ee
with different statistical operators
\be
\label{73}
 \hat\rho_\nu \equiv
\frac{\exp(-\bt H_\nu)}{{\rm Tr}_{\cH_\nu}\exp(-\bt H_\nu)}
\ee
and over different spaces of microscopic states typical of the given phase.
The phase with broken gauge symmetry corresponds to the nonzero average
\be
\label{74}
 \lgl S_j^z \rgl_1 \neq 0 \;  ,
\ee
while the phase with the conserved symmetry gives
\be
\label{75}
 \lgl S_j^z \rgl_2 = 0 \;  .
\ee

The order parameters are defined as
\be
\label{76}
 s_\nu \equiv \frac{2}{N} \sum_{j=1}^N  \lgl S_j^z \rgl_\nu .
\ee
For the phases with the broken symmetry and unbroken symmetry, one has,
respectively,
\be
\label{77}
 s_1 \neq 0 \; , \qquad s_2 = 0 \;  .
\ee

To calculate the order parameter $s_1$ and the phase probabilities $w_\nu$,
we resort to the mean-field approximation, with the use of the notations
\be
\label{78}
u \equiv \frac{A}{I+B} \; , \qquad b \equiv \frac{B}{I+B} \; , \qquad
\om \equiv \frac{\Om_0}{I+B} \; ,
$$
$$
A  \equiv \frac{1}{N} \sum_{i\neq j}^N A_{ij} \; , \qquad
B  \equiv \frac{1}{N} \sum_{i\neq j}^N B_{ij} \; , \qquad
I  \equiv \frac{1}{N} \sum_{i\neq j}^N I_{ij} \; .
\ee

Then, the minimization of the grand potential $\Omega$ yields the equations
$$
w_1 = \frac{2u+\om_1 x_1 -\om_2 x_2}{4u-(1-b)s_1^2} \; , \qquad
w_2 = 1 - w_1 \;  ,
$$
\be
\label{79}
\sqrt{(w_1s_1)^2+\om^2} = w_1 \tanh\left [
\frac{(1-b)w_1}{2T} \; \sqrt{(w_1s_1)^2+\om^2} \right ] \; ,
\ee
in which the effective temperature $T$ is measured in units of $I + B$ and
$$
 \om_1 = (1-b)\om \; , \qquad \om_2 = \om - bw_2 x_2 \; ,
$$
\be
\label{80}
x_1 = \frac{\om}{w_1} \; , \qquad
x_2 = \tanh\left ( \frac{w_2 \omega_2}{2T} \right ) \; .
\ee

We solved these equations numerically for $b \ll 1$, $\omega \ll 1$, and
for varying $u$. At $T = 0$, the order parameter $s_1 = 1$. It monotonically
decreases with rising $T$ up to a phase-transition point, where $s_1$ becomes
zero and the system transfers to a disordered phase with unbroken gauge
symmetry. The phase transition is of second order for $u \leq 0$ and
$u \geq 3/2$, while of first order for $0 < u < 3/2$. The behavior of $w_1$
and $s_1$ as functions of $T$ is shown in Figs. 1 and 2. This model
illustrates that, for some system parameters and for sufficiently high
effective temperature that includes the injected energy, the system, can,
first, become a mixture of two phases, one with broken gauge symmetry and
the other with unbroken symmetry, and at a large effective temperature $T$ 
the whole system transfers to the disordered phase with unbroken symmetry,
that is, becomes normal gauge-symmetric phase.

\section{Conclusion}

The idea is advanced that strong perturbations of an initially equilibrium 
Bose-condensed gas lead to the sequence of nonequilibrium states whose order 
is inverse to the sequence of states arising in the process of the Bose-gas 
relaxation from an initial nonequilibrium non-condensed state to its 
equilibrium Bose-condensed state. 

We have described a general approach for constructing statistical models of 
nonequilibrium Bose gases. The method is based on the averaging over 
heterogenous configurations of a nonequilibrium system. 

A statistical model of grain turbulence is suggested, whose general properties 
are formulated. Numerical calculations for a simple model, in the frame of a 
mean-field approximation, show that the grain turbulence can exist in a region 
of injected energies. Increasing the amount of energy, injected into the system 
above the threshold, leads to the destruction of the regime of grain turbulence, 
so that the system passes to a normal non-condensed state with the preserved 
gauge symmetry. The gauge-symmetric phase represents the wave turbulence of 
normal, non-condensed gas. The transformation of the gauge-broken phase into 
the gauge-symmetric phase plays the role of the transition from the grain 
turbulence to wave turbulence.

\vskip 5mm

{\bf Acknowledgements}

\vskip 2mm

We are grateful for discussions to V.S. Bagnato, R.G. Hulet, and A.N. Novikov. 
Financial support from the Russian Foundation for Basic Research 
(grant $\# 14-02-00723$) is appreciated.

\newpage

\begin{center}
{\bf{\Large Figure Captions}}
\end{center}

\vskip 2cm

{\bf Fig. 1}. Probability of the ordered phase as a function of the dimensionless
effective temperature for different parameters $u$: (1) $u=0$; (2) $u=0.1$;
(3) $u=0.3$; (4) $u=0.4$; (5) $u=0.51$; (6) $u=0.75$; (7) $u=1$; (8) $u=1.5$.
The dots mark the points of the transition from the broken-symmetry phase 
with a nonzero order parameter to the gauge-symmetric phase with zero order
parameter.

\vskip 2cm

{\bf Fig. 2}. Order parameter as a function of the dimensionless
effective temperature for different parameters $u$: (1) $u=0$; (2) $u=0.1$;
(3) $u=0.3$; (4) $u=0.4$; (5) $u=0.51$; (6) $u=0.75$; (7) $u=1$; (8) $u=1.5$.
The dots mark the points of the transition from the broken-symmetry phase 
with a nonzero order parameter to the gauge-symmetric phase with zero order
parameter.

\newpage

\begin{figure}[ht]
\centerline{\includegraphics[width=16cm]{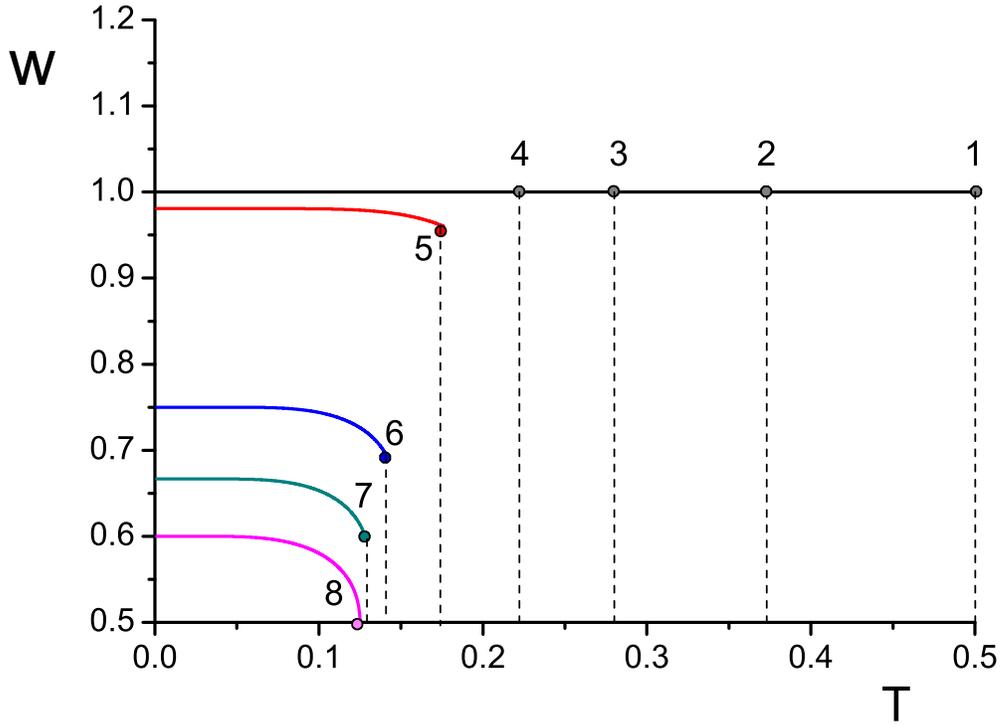}}
\caption{Probability of the ordered phase as a function of the dimensionless
effective temperature for different parameters $u$: (1) $u=0$; (2) $u=0.1$;
(3) $u=0.3$; (4) $u=0.4$; (5) $u=0.51$; (6) $u=0.75$; (7) $u=1$; (8) $u=1.5$.
The dots mark the points of the transition from the broken-symmetry phase 
with a nonzero order parameter to the gauge-symmetric phase with zero order
parameter.
}
\label{fig:Fig.1}
\end{figure}

\begin{figure}[ht]
\centerline{\includegraphics[width=16cm]{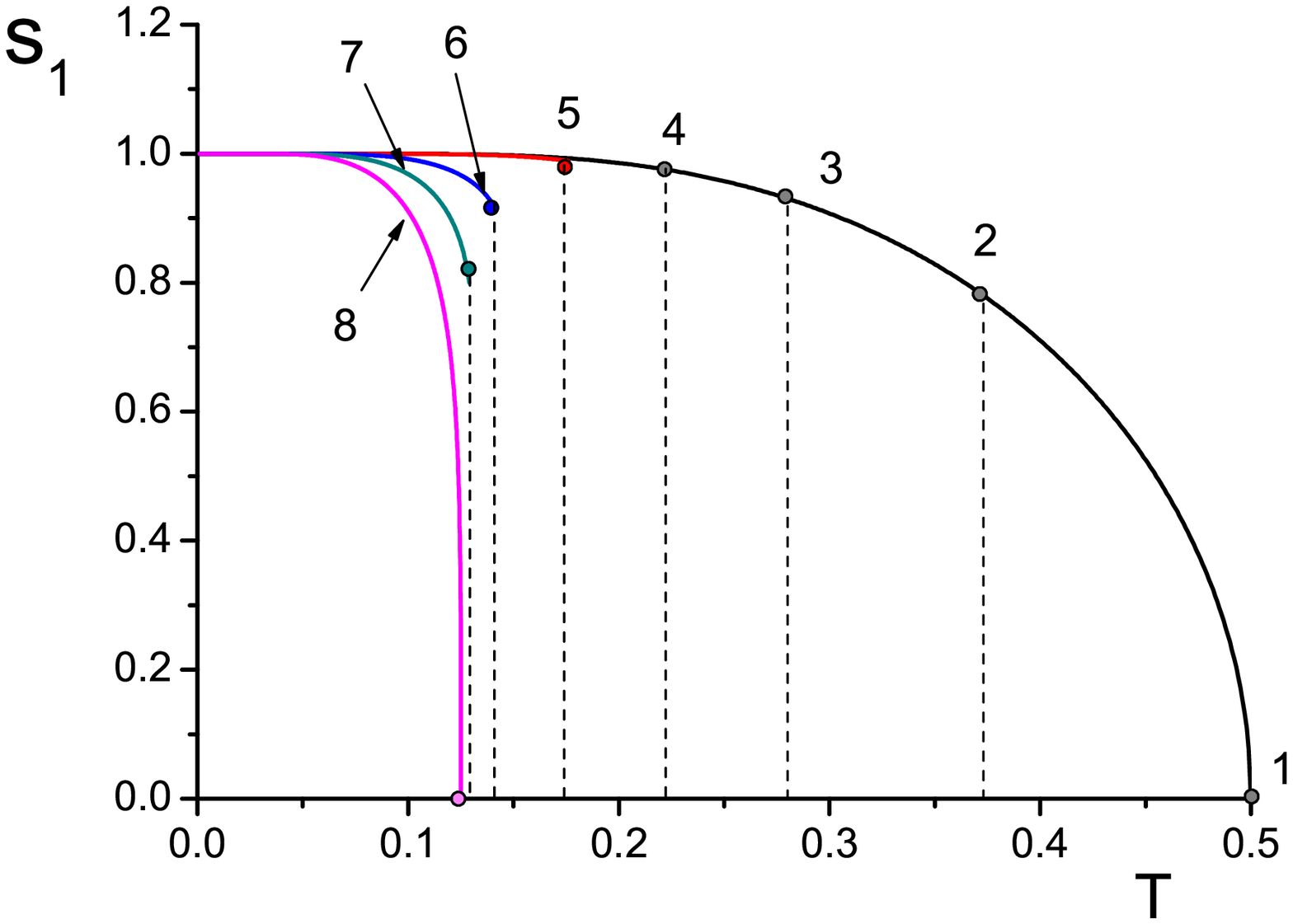}}
\caption{Order parameter as a function of the dimensionless
effective temperature for different parameters $u$: (1) $u=0$; (2) $u=0.1$;
(3) $u=0.3$; (4) $u=0.4$; (5) $u=0.51$; (6) $u=0.75$; (7) $u=1$; (8) $u=1.5$.
The dots mark the points of the transition from the broken-symmetry phase 
with a nonzero order parameter to the gauge-symmetric phase with zero order
parameter.
}
\label{fig:Fig.2}
\end{figure}

\end{document}